\documentclass[11pt]{article}
\usepackage{amssymb}
\usepackage{amsmath}

\def\( {\left(}
\def\) {\right)}
\def\[ {\left[}
\def\] {\right]}
\newcommand{\beq}{\begin{equation}}
\newcommand{\eeq}{\end{equation}}
\newcommand{\beqn}{\begin{eqnarray}}
\newcommand{\eeqn}{\end{eqnarray}}

\newcommand{\om}{\omega}

\newcommand{\si}{\sigma}

\newcommand{\te}{\theta}

\newcommand{\pa}{\partial}
\newcommand{\al}{\alpha}

\newcommand{\lga}{\longrightarrow}

\newcommand{\NP}[1]{ {\it Nucl.~Phys.} {\bf #1}}

\newcommand{\PR}[1]{ {\it Phys.~Rev.} {\bf #1}}
\newcommand{\PRL}[1]{ {\it Phys.~Rev.~Lett.} {\bf #1}}

\begin{document}
\begin{titlepage}
\setcounter{page}{1}
\renewcommand{\thefootnote}{\fnsymbol{footnote}}

\begin{flushright}
ucd-tpg:09-01\\
arXiv:0904.0587
\end{flushright}

\vspace{5mm}
\begin{center}

{\Large\bf Confined Dirac Fermions in a Constant Magnetic Field}

\vspace{5mm}

{\bf Ahmed Jellal$^{a}$\footnote{ajellal@ictp.it, jellal@ucd.ac.ma}},
{\bf Abdulaziz D. Alhaidari$^{b}$\footnote{haidari@sctp.org.sa}}
and 
{\bf Hocine Bahlouli$^{c}$\footnote{bahlouli@kfupm.edu.sa}}
%and {\bf Ahmed Jellal$^{c}$\footnote{ajellal@ictp.it, jellal@ucd.ac.ma}}

\vspace{5mm}

%{\em $^{a}$The Abdus Salam International Centre for Theoretical Physics},\\
%{\em  Strada Costiera 11, 34014 Trieste, Italy}\\

%{\em and} \\

{ \em $^{a}$Theoretical Physics Group, % Department of Physics,
Faculty of Sciences,  Choua\"ib Doukkali University},\\
{\em Ibn Ma\^achou Road, PO Box 20, 24000 El Jadida,
Morocco}\\

{\em $^{b}${%\footnotesize 
Saudi Center for Theoretical Physics, Dhahran 31261, Saudi Arabia}}

{\em $^{c}${%\footnotesize 
Physics Department, King Fahd University
of Petroleum $\&$ Minerals,\\
 Dhahran 31261, Saudi Arabia}}
%\vspace{30mm}

%A. Jellal$^{a}$, A. D. Alhaidari$^{b,c}$, and H. Bahlouli$^{c}$
%\end{indentation}
%\end{center}

\vspace{30mm}

\begin{abstract}

We obtain an exact solution of the Dirac equation in $(2+1)$-dimensions 
in the presence of a constant magnetic field normal to the
plane together with a two-dimensional Dirac-oscillator potential coupling. The
solution space consists of a positive and negative energy solution,
each of which splits into two disconnected subspaces depending on the
sign of an azimuthal quantum number, $k  = 0, \pm 1, \pm
2,\cdots$ and whether the cyclotron frequency
is larger or smaller than the oscillator frequency. 
The spinor wavefunction is written in terms of
the associated Laguerre polynomials. For negative
$k$, the relativistic energy
spectrum is infinitely degenerate due to the fact that it is
independent of $k$. 
We compare our results with already published work and point out
the relevance of these findings to a systematic formulation of the
relativistic quantum Hall effect in a confining potential.
\end{abstract}

%\vspace{2mm}
\begin{flushleft}
\end{flushleft}

\end{center}
\end{titlepage}

%\newpage

%%%%%%%%%%%%%%%%%%%%%%%%%%%%%%%%%%%%%%%%%
\section{Introduction}
%%%%%%%%%%%%%%%%%%%%%%%%%%%%%%%%%%%%%%%%%%%%%

Due to recent technological advances in nano-fabrication there were a 
lot of interest in the study of low dimensional quantum systems such as quantum 
wells, quantum wires and quantum dots~\cite{ref1}. In particular, there has been 
considerable amount of work in recent years on semiconductors confined structures, 
which finds applications in electronic and optoelectronic devices. The application of 
a magnetic field perpendicular to the hetero-structure plane quantizes the energy levels 
in the plane, drastically affecting the density of states giving rise to the famous 
quantum Hall effect (QHE)~\cite{prange}. The latter remains among the most interesting 
phenomena observed in physics because of its link to different theories and subjects. 

The stationary state associated with the motion of electrons in a uniform magnetic 
field is a well-known textbook problem \cite{ref3}. 
It results in a sequence of the quantized Landau energy levels and associated wavefunctions characterizing the dynamics in the two-dimensional (2D) plane normal to the applied magnetic field. This quantization has important consequences in condensed matter physics ranging from the classical de Hass-van Alphen effect in metals
\cite{ref4}
%[4]
to QHE in semiconductors~\cite{prange}. 
The relativistic extension of these models turned out to be of great importance in the description of 
2D quantum phenomena such as QHE in graphene
\cite{ando, sharapov, novoselov, zhang}. 
In fact, several condensed matter phenomena point out to the existence of a
 $(2+1)$-dimensional energy spectrum determined by the
relativistic Dirac equation~\cite{ref6}. For very recent works,
one may consult references~\cite{goerbig,toke,khvesh,toke2,jellal2}. 

%{I still plan to write more later.}
Motivated by different investigations on the Dirac fermions in $(2+1)$-dimensions, we give an exact solution of a problem that has been studied at various levels by researchers dealing with different physical phenomena. We have done so by considering a relativistic particle subjected to an external magnetic field as well as to a confining potential. By introducing a similarity transformation, we show that the system can be diagonalized in a simple way. Solving the eigenvalue equation, we end up accounting for the full space of the eigenfunctions that includes all cases related to different physical settings. More precisely, from the nature of the problem we get separate angular and radial solutions. The radial equation leads to a "kinetic balance" relation between the two-spinor components. In fact, depending on the range of values of three physical quantities, the full solution space splits into eight disconnected subspaces as summarized in Table 1 in Sec. 2.2. This allowed us to obtain various solutions and emphasis similarities to, and differences from, already published work elsewhere~\cite{villalba}.

On the other hand, we give discussions of our results based on different physical settings. In fact, we show that for week and strong magnetic field there is a symmetry that allows us to go from positive to negative energy solutions (states and spectrum). This can be done by interchanging the confinement frequency $\om$ with the cyclotron’s $\om_c$ and vise versa. This suggests defining an effective magnetic field that produces the effective quantized Landau levels. In both cases, there is a degeneracy of the Landau levels where each quantum number $n$ is $k$-times degenerate; in analogy with the non-relativistic case~\cite{jellal1}. 
For the intermediate magnetic field case, it is underlined that the degeneracy is possible. Finally, 
we compare our findings with those in a very significant work by Villalba and Maggiolo~\cite{villalba}.
The full rich space of solutions suggested enabled us to carry out a deeper analysis in relation to various physical quantities. For instance, we obtained, as expected in the absence of applied voltage, a null current density for both directions in the Cartesian representation. However, this is not the case in polar coordinate. In fact, we show that the radial current vanishes whereas the angular component does not. It is dependent on various physical parameters in the problem. These results are summarized in Table 3 showing clearly the dependence of these values on the given subspace. This may offer an alternative approach for a systematic study and understanding of the anomalous QHE~\cite{novoselov,zhang}. 
Additionally, we discuss the non-relativistic limit of the problem.

The paper is organized as follows. In section $2$, we give the theoretical 
formulation of the problem where a similarity transformation
is introduced to simplify the process for obtaining
the solutions (spinor wavefunction and energy spectrum).  
We use the "kinetic balance" relation to obtain a second order differential equation
for one of the two-spinor components. 
The second spinor component is obtained from this using the "kinetic balance" relation.
The relativistic energy eigenvalues and corresponding spinor wavefunctions are 
obtained as elements in the eight subspaces of the full and complete Hilbert space. 
In section $3$, we discuss the physical meaning of our results and their potential 
application to QHE. To analyze the transport properties of the 
system we determine the current density in section $4$ and the non-relativistic case 
in section $5$. Finally, we conclude by discussing the main results and possible 
extension of our work.

%%%%%%%%%%%%%%%%%%%%%%%%%%%%%%%%%%%%%%%%%%%%%%%%%%%%%%%%
\section{Formulation of the problem}
%%%%%%%%%%%%%%%%%%%%%%%%%%%%%%%%%%%%%%%%%%%%%%%%%%%%%%%%%

We start by formulating the problem in terms of our language~\cite{alhaidari}. This is done 
by introducing a similarity transformation of the Dirac equation in polar coordinates. 
This will be convenient to handle the "kinetic balance" relation and therefore derive the 
full spectrum as a complete Hilbert space.

%%%%%%%%%%%%%%%%%%%%%%%%%%%%%%%%%%%%%%%%%%%%%%%%%%%%%%%%
\subsection{Hamiltonian system}
%%%%%%%%%%%%%%%%%%%%%%%%%%%%%%%%%%%%%%%%%%%%%%%%%%%%%%%%%

The problem of a charged particle moving in a constant
magnetic field $\vec B = B\,\hat z$ is a 2D problem 
in the plane normal to the field [the Cartesian
$(x,y)$-plane or cylindrical
$(r,\te)$-plane]. In the
relativistic units, $\hbar = c = 1$, the Dirac equation in $(2+1)$-dimensions for a
spinor of charge $e$ and mass $m$ in the electromagnetic potential
${A_\mu } = ({A_0},\vec A)$ reads as follows
\begin{equation}
\left[ {{{i}}{\gamma ^\mu }({\partial _\mu } +
{{i}}e{A_\mu }) - m} \right]\psi  = 0\label{eq1}, \qquad \mu =0,1,2
\end{equation}
where the summation convention over repeated indices is used.
${\gamma ^\mu } = \left( {{\gamma ^0},\vec \gamma }
\right)$ are three unimodular square matrices satisfying the
anti-commutation relation: 
\begin{equation}
\left\{ {{\gamma ^\mu
},{\gamma ^\nu }} \right\} = {\gamma ^\mu }{\gamma ^\nu }
 + {\gamma ^\nu }{\gamma ^\mu } = 2{{\cal G}^{\mu
\nu }}
\end{equation}
 where ${\cal G}$ is the metric of
Minkowski space-time, which is equal to $\mbox{diag} ( + \,\, - \,\,- )$. 
A minimal irreducible matrix representation that
satisfies this relation is taken as 
${\gamma ^0} =
{\sigma _3}$, $  \vec \gamma  = {{i}}\,\vec
\sigma$
 where $\left\{ {{\sigma _i}} \right\}_{i= 1}^3$ 
are the $2\times 2$ hermitian Pauli spin matrices:
\begin{equation}
{\sigma _1} = \left( {\begin{array}{*{20}{c}}
0 & 1  \\
1 & 0  \\
\end{array}} \right),
\qquad
{\sigma _2} = \left( {\begin{array}{*{20}{c}}
0 & -i  \\
i & 0  \\
\end{array}} \right),\qquad
{\sigma _3} = \left( {\begin{array}{*{20}{c}}
1 & 0  \\
0 & -1  \\
\end{array}} \right).
\end{equation}
Equation (\ref{eq1}) could be rewritten as
\begin{equation}
{{i}}{{\partial  \over {\partial t}}}\psi
 = \left( { - {{i}}\,\vec \alpha  \cdot \vec \nabla  + e\vec \alpha
\cdot \vec A + e{A_0} + m\beta } \right)\psi \label{eq3}
\end{equation}
where $\vec \alpha $ and $\beta$ are the hermitian matrices:
$\vec \alpha  = {{i}}\,{\sigma _3}\vec \sigma, \beta  = {\sigma _3}$. 
We will
see below that the symmetry of the problem is preserved even if we
introduce an additional coupling to the 2D Dirac-oscillator potential.
This coupling is introduced by the substitution 
$
\vec \nabla
 \to \vec \nabla  + m\omega \vec r\beta$
where $\omega$ is the oscillator frequency.
For time independent potentials, the two-component spinor wavefunction
$\psi (t,r,\theta )$ is written as 
\begin{equation}
\psi (t,r,\theta ) = {e^{- {{i}}\varepsilon t}}\psi (r,\theta )
\end{equation} 
and (\ref{eq3})
%(2.3)
becomes the energy eigenvalue wave equation $
({\cal H} - \varepsilon )\psi  = 0$
 where $\varepsilon$ is the relativistic energy.
The Dirac Hamiltonian ${\cal H}$ is the
$2\times 2$ matrix operator:
\begin{equation}
{\cal H} = {{\cal H}_0} + {{i}}\,{\sigma _3}\vec \sigma  \cdot \hat
r\,{{\cal H}_r} + {{i}}\,{\sigma _3}\vec \sigma  \cdot \hat \theta
\,{{\cal H}_\theta }
\label{eq4}
\end{equation}
where ($\hat r,\hat \theta $) are the unit vectors
in cylindrical coordinates and 
%\begin{equation}
\beqn
{{\cal H}_0} &=& e{A_0} + m{\sigma _3}, \nonumber \\ 
{{\cal H}_r} &=& -i\pa_r + eA_r -i m\om r\sigma_3, 
\\
{{\cal H}_{\theta}} &=& -{i\over r} \pa_{\theta} + eA_{\theta}. \nonumber 
\label{eq5}
%(2.5)
\eeqn
%\end{equation}
For regular solutions of (\ref{eq3}), square integrability (with respect to the measure
${d^2}\vec r = r\,dr\,d\theta $) and the boundary
conditions require that $\psi (r,\theta )$
satisfies
\begin{equation}
{\left. {\sqrt r \,\psi (r,\theta )} \right|_{\scriptstyle r = 0
\hfill \atop
\scriptstyle r \to \infty  \hfill}} = 0, \qquad
\psi (\theta+2\pi) = \psi (\theta).
\label{eq6}
\end{equation}

To simplify the construction of the solution, we look for a local
$2\times2$ similarity transformation
$\Lambda (r,\theta )$ that maps the cylindrical
projection of the Pauli matrices $(\vec \sigma  \cdot \hat
r$, $\vec \sigma  \cdot \hat \theta )$ into
their canonical Cartesian representation $({\sigma
_1}, {\sigma _2})$, respectively~\cite{explain}. That is
\begin{equation}
\Lambda\, \vec \sigma  \cdot \hat r\,{\Lambda ^{ - 1}} = {\sigma _1},
\qquad
\Lambda \,\vec \sigma  \cdot \hat \theta\,{\Lambda ^{ - 1}} = {\sigma _2}.
\label{eq7}
\end{equation}
A $2\times 2$ matrix that satisfies this
requirement is
\begin{equation}
\Lambda (r,\theta ) = \lambda (r,\theta )\,{e^{{{}{{{i}}
\over 2}}{\sigma _3}\theta }} %,\qquad
\label{eq8}
\end{equation}
where $\lambda (r,\theta )$ is a
$1\times 1$ real function and the exponential is a
$2\times 2$ unitary matrix. The Dirac Hamiltonian
(\ref{eq4})
%(2.4)
gets mapped into
\begin{equation}H = \Lambda {\cal H}{\Lambda ^{ - 1}} = {H_0} -
{\sigma _2}{H_r} + {\sigma _1}{H_\theta }\label{eq9}
%(2.8)
\end{equation}
where different operators are given by
\beqn
{H_0} &=& {{\cal H}_0}, \nonumber\\
H_r &=&-i\left(\partial _r - {\lambda _r \over\lambda }\right) + i e{A_r} - im\omega r\sigma_3, \\
H_{\theta} &=& -
{i \over r}\left(\partial _\theta  -
{\lambda _\theta \over \lambda}  -{i\over 2}\sigma_3 \right)+ eA_\theta\nonumber
\label{eq10}
%(2.9)
\eeqn
with  ${\lambda _k} = {\partial _k}\lambda $.
Therefore, the 2$\times{}$2 Dirac Hamiltonian
becomes
%\begin{equation}
%H=
%\small{
%\left( {\begin{array}{*{20}{c}}
%{m + e{A_0}} & {{\partial _r} - {{{{\lambda _r}} \over
%\lambda }} + {{1 \over {2r}}} - {{i}}e{A_r} + m\omega r -
%{{{{i}} \over r}}\left( {{\partial _\theta } +
%{{{{\lambda _\theta }} \over \lambda }}} \right) + e{A_\theta
%}}  \\
%{ - {\partial _r} + {{{{\lambda _r}} \over \lambda }} -
%{{1 \over {2r}}} + {{i}}e{A_r} + m\omega r -
%{{{{i}} \over r}}\left( {{\partial _\theta } +
%{{{{\lambda _\theta }} \over \lambda }}} \right) + e{A_\theta
%}} & { - m + e{A_0}}  \\
%\end{array}} \right)}\label{eq11}
%\end{equation}
%\begin{equation}
 %h= {a\choose b \\ c\choose d}
%\end{equation}
\begin{equation}
\label{smham}
H =
\small
\begin{pmatrix}
 m + eA_0  & \partial _r - {\lambda _r \over\lambda } + 
{1 \over 2r} + i e A_r - m\omega r - {i \over r}\left( \partial_\theta  -
{\lambda _\theta  \over \lambda } \right) + e A_\theta \\
-\partial _r + {\lambda _r\over \lambda } -
{1 \over 2r} - i e A_r + m\omega r - {i \over r} \left(\partial _\theta  -
{\lambda _\theta \over \lambda } \right) + e A_\theta
   & - m + e A_0
\end{pmatrix}.
\end{equation}
Thus, hermiticity of (\ref{smham}) requires that
\begin{equation}
{\lambda _\theta } = 0, \qquad
{{{\lambda _r}} \over
\lambda } - {{1 \over {2r}}}=0
\label{eq12}
%(2.11)
\end{equation}
and fixes the exact form of the modulus of similarity transformation
to be
$\lambda (r,\theta ) = \sqrt r$. 
It is
interesting to note that ${\lambda ^2}$ turns out
to be the integration measure in 2D cylindrical coordinates. We could
have eliminated the $\lambda$ factor in
the definition of $\Lambda$ in (\ref{eq8})
%(2.7)
by proposing that the new spinor wavefunction
{$\chi$} be replaced by
${{1 \over {\sqrt r }}}\chi (r,\theta )$.
In that case, the transformation matrix $\Lambda{}$
becomes simply ${e^{{{}{{{i}} \over 2}}{\sigma _3}\theta }}$, which is unitary. However, making
the presentation as above gave us a good opportunity to show (in a
different approach) why is it customarily to take the radial component
of the wavefunction in 2D cylindrical coordinates to be proportional to
${1 \over \sqrt r }$. 
Finally, we obtain
the $(2+1)$-dimensional Dirac equation
$
\left( {H - \varepsilon }
\right)\chi  = 0$
for a charged spinor in static
electromagnetic potential as
\begin{equation}
\left( {\begin{array}{*{20}{c}}
{m + e{A_0} - \varepsilon } & {{\partial _r} + {{i}}e{A_r} -
m\omega r - {{{{i}} \over r}}{\partial _\theta } +
e{A_\theta }}  \\
{ - {\partial _r} - {{i}}e{A_r} - m\omega r -
{{}{{{i}} \over r}}{\partial _\theta } + e{A_\theta }} & { -
m + e{A_0} - \varepsilon }  \\
\end{array}} \right)\left( {\begin{array}{*{20}{c}}
{\mathop {{\chi _ + }(r,\theta )}\limits_{} }  \\
{{\chi _ - }(r,\theta )}  \\
\end{array}} \right) = 0\label{eq13}
%(2.12)
\end{equation}
where ${\chi_\pm }$ are the components of the
transformed wavefunction $\left| \chi  \right\rangle  =
\Lambda \left| \psi  \right\rangle $.
This equation will be solved by choosing an appropriate gauge
to end up with the full Hilbert space.

%%%%%%%%%%%%%%%%%%%%%%%%%%%%%%%%%%%%%%%%%%%%%%%%%%%%%%%%%%%%
\subsection{Eigenvalues and wavefunctions}
%%%%%%%%%%%%%%%%%%%%%%%%%%%%%%%%%%%%%%%%%%%%%%%%%%%%%%%%%%%%

Now, we specialize to the case where a constant magnetic
field of strength $B$ is applied at right angles to
the ($r, \theta $)-plane, which is
$\vec B = B\,\hat z$. Therefore, the
electromagnetic potential has the time and space components:
\begin{equation}
{A_0} = 0, \qquad \vec A(r,\theta ) =
{{1 \over 2}}Br\,\hat \theta. 
\end{equation} 
Consequently, (\ref{eq13})
%(2.12)
becomes completely separable and we can write the spinor wavefunction as 
\begin{equation}
{\chi _ \pm }(r,\theta ) = {\phi _ \pm
}(r)\,\tau (\theta ).
\end{equation} 
Thus, the angular component satisfies
$
- {{i}}{{}{{d\tau } \over {d\theta }}} = \xi
\,\tau 
$ 
where {$\xi{}$} is a real
separation constant giving the function: 
\begin{equation}
\tau (\theta ) = {{1
\over {\sqrt {2\pi } }}}{e^{{{i}}\xi \theta }}.
\end{equation} 
On the other hand, the boundary condition $\psi (\theta  + 2\pi ) = \psi (\theta)$  
requires that 
%\begin{equation}
${e^{{{i}}\,2\pi \xi }}{e^{ -
{{i}}{\sigma _3}\pi }} =  + 1$ 
which, in turn, demand that
$
{e^{{{i}}\,2\pi \xi }} =  - 1
$ 
giving the following quantum number:
\begin{equation}
\xi  = {{1 \over 2}}\kappa, \qquad
\kappa=\pm 1, \pm 3, \pm 5 \cdots.
\label{eq14}
%(2.13)
\end{equation}
Consequently, the Dirac equation for the two-component radial spinor
is reduced to
\begin{equation}\left( {\begin{array}{*{20}{c}}
{m - \varepsilon } & {{{d \over {dr}}} +
{{\xi  \over {r}}} + Gr}  \\
{ - {{d \over {dr}}} + {{\xi  \over {r}}} +
Gr} & { - m - \varepsilon }  \\
\end{array}} \right)\left( {\begin{array}{*{20}{c}}
{\mathop {{\phi _ + }(r)}\limits_{} }  \\
{{\phi _ - }(r)}  \\
\end{array}} \right) = 0\label{eq15}
%(2.14)
\end{equation}
where the physical constant $G$ is given by
$
G = m\left(\omega_c  - \om\right)$ 
and $\om_c$ is the cyclotron frequency $\om_c={{eB \over 2m}}$.
Thus, the presence of the 2D Dirac-oscillator
coupling did, in fact, maintain the symmetry of the problem as stated
below (\ref{eq3}). Moreover, its introduction is equivalent to 
changing the magnetic
field as 
$eB \lga eB - 2m\omega$.
 As a result of
the wave equation (\ref{eq15}), the two spinor components satisfy 
the "kinetic balance" relation:
\begin{equation}
{\phi _ \mp }(r) = \frac{1}{{\varepsilon  \pm m}}\left[ { \mp
\frac{d}{{dr}} + \frac{\xi} {r} + Gr} \right]{\phi _ \pm }(r)
\label{eq16}
%(2.15)
\end{equation}
where $\varepsilon  \ne  \pm m$. Therefore, the
solution of the problem with the top/bottom sign corresponds to the
positive/negative energy solution. Using the "kinetic balance" relation
(\ref{eq16})
%(2.15)
to eliminate one component in terms of the other in (\ref{eq15})
results in the following Schr\"{o}dinger-like differential equation
for each spinor component: 
\begin{equation}
\left\{ { - \frac{{{d^2}}}{{d{r^2}}} + \frac{{{{}{\xi}}\left( {{{}{\xi}} \mp 1}
\right)}}{{{r^2}}} + {G^2}{r^2} + \left[ {{m^2} - {\varepsilon ^2} +
G\left( {2\xi  \pm 1} \right)} \right]} \right\}{\phi _ \pm }(r) = 0.
\label{eq17}
\end{equation}
Again, we stress that this equation gives \textbf{only one} radial
spinor component. One must choose either the top or bottom sign to
obtain the component that corresponds to the positive or negative
energy solutions, respectively. The second component is obtained by
substituting this into the "kinetic balance" relation (\ref{eq16}). 
Nonetheless, we only need to find one solution (the positive- or
negative-energy solution), because the other is obtained by a simple
map. For example, the following map takes the positive energy solution
into the negative energy solution:
\begin{equation}
\varepsilon  \lga  - \varepsilon,\qquad \kappa  \lga  - \kappa,\qquad G \lga
- G,\qquad {\phi _ \pm } \lga {\phi _ \mp }
\label{eq18}
%(2.17)
\end{equation}
which, in fact, is the $\cal{CPT}$ transformation. Here the charge
conjugation $C$ means that 
$e \lga  - e$ and  $\omega\lga-\omega$
or the exchange of $\om$ and $\om_c$.
It is easy to check that the above map (\ref{eq18}) originates from 
the fact that the Dirac equation (\ref{eq15}) is invariant under such 
transformation. Hence, we just need to solve for positive energies and use 
the above transformation to obtain the negative energy solutions. The total 
spinor wavefunction reads as follows
\begin{equation}
\psi (r,\theta ) = \frac{1}{{\sqrt r }}\,{e^{{{}{{{i}}
}}\xi \theta }} {e^{ - {{}{{{i}} \over 2}}{\sigma
_3}\theta }} \phi (r)
\label{eq19}
%(2.18)
\end{equation}
where $\phi (r)$ has two components, such as
\begin{equation}
\phi= \left( {\begin{array}{*{20}{c}}
 \phi _ + \\
 \phi _ -\\
\end{array}} \right).
%\phi  = \left( {_{{\phi _ - }}^{{\phi _ + }}}
%\right).
\end{equation} 
Equation (\ref{eq17})
looks like the non-relativistic oscillator problem with a certain
parameter map of the frequency, angular momentum, and energy. For
regular solutions of (\ref{eq17}), 
the bound states %(where $\left| \varepsilon  \right| < m$) 
will be of the form
\begin{equation}
{\phi _ \pm }\sim 
 {z^\mu }{e^{ - {z \mathord{\left/
{\vphantom {z 2}} \right.
\kern-\nulldelimiterspace} 2}}}L_n^\nu (z)
\label{eq20}
%(2.19)
\end{equation}
where $L_n^\nu (z)$ is the associated Laguerre
polynomial of order $n = 0,1,2,\cdots$ and
$z = {\rho ^2}{r^2}$. The constants
$\left\{ {\mu ,\nu ,\rho } \right\}$ are real and
related to the physical parameters $B$,
{$\omega$} and {$\xi$}. Square integrability and the
boundary conditions require that $2\mu  \ge {{}{1
\over 2}}$ and $\nu  >  - 1$.
%\end{indentation}

Substituting the ansatz (\ref{eq20})
%(2.19)
into (\ref{eq17})
%(2.16)
and using the differential equation for the Laguerre polynomial shown
in Appendix A, we obtain four equations. Three of them determine the
parameters $\left\{ {\mu ,\nu ,\rho } \right\}$ and
one determines the energy spectrum. The first three are
\beqn
2\mu &=& \nu + {{1 \over 2}}, \qquad
\rho^2=G, \nonumber\\
\nu  &=& \pm \left\{ {\begin{array}{*{20}{c}}
\xi -{1\over 2 }, \qquad  { \varepsilon > 0}   \\
\xi +{1\over 2}, \qquad
{\varepsilon < 0}.  \\
\end{array}}
\right.
\label{eq21}
%(2.20)
\eeqn
For regular solutions of (\ref{eq17}), the $\pm $ sign in the expression for $\nu $ 
corresponds to $\pm \xi > 0$. Now, the fourth equation gives the 
following (positive and negative) energy spectra:
\begin{equation}\label{2.22}
\varepsilon _{n,\xi}^{\pm}  = \pm m\sqrt{1 + {2|G|\over m^2} 
\left[2n+1 \pm {s-s'\over 2} +\xi (s+s')\right]}
\end{equation}
where $s=\mbox{sign}\left (G \right)={|G|\over G}$ and $s'=\mbox{sign}\left (\xi \right)$. 
The sign of $G$ depends on 
whether the oscillator frequency $\omega $ is larger or smaller than the cyclotron 
frequency {$\omega $}$_{c}$.
To compare our work with frequently used notation in the literature,
we can replace the quantum number
{$\xi$} by $k +
{1\over 2}$, where $k = 0, \pm 1, \pm 2,\cdots $
and $\xi \lga -\xi$ imply that $k \lga -k-1$. 
In that case, one may write the energy spectrum as
positive eigenvalues:
\begin{equation}\label{2.22a}
\varepsilon _{n,k}^{+}  = m\sqrt{1 + {2|G|\over m^2} 
\left[2n +1 +s +{k} (s+s')\right]}
\end{equation}
and negative ones:
\begin{equation}\label{2.22b}
\varepsilon _{n,k}^{-}  = - m\sqrt{1 + {2|G|\over m^2} 
\left[2n +1+s' +{k} (s+s')\right]}
\end{equation}
where $s'= +1$ for $k$ = 0.
It is interesting to note that for $\xi G  < 0$
the spectrum is infinitely degenerate because it is independent of
$\xi$. However, for
$ \xi G > 0$ the degeneracy is finite and equal to
$n + k + 1$. Substituting the wavefunction
parameters given by (\ref{eq21})
into the ansatz (\ref{eq20})
gives for $\varepsilon  > 0$:
\begin{equation}
{\phi _ + }(r) =  {x^{\left| k +{1\over 2}  \right|}} \, e^{ - {1 \over 2}x^2}
\\
 \left\{ {\begin{array}{*{20}{c}}
A_{n,k}^{++}\, L_n^{k}\left(x^2\right),\qquad~~~ &  {k  \geq 0}  \\
A_{n,k}^{+-} \, x
L_n^{-k} \left(x^2\right), \qquad & {k  < 0}  \\
\end{array}} \right. 
%\end{array}
\label{eq22}
%(2.26)
\end{equation}
as well as for $\varepsilon  < 0$:
\begin{equation}
%\varepsilon  < 0:\qquad
{\phi _ - }(r) =  {x^{\left| k +{1\over 2}  \right|}} \, e^{ - {1 \over 2}x^2}
\\
 \left\{ {\begin{array}{*{20}{c}}
A_{n,k}^{-+}\, x L_n^{k+1}\left(x^2\right),\qquad &  {k  \geq 0}  \\
A_{n,k}^{--} \,
L_n^{-k-1} \left(x^2\right), \qquad & {k  < 0}  \\
\end{array}} \right. 
\label{eq23}
%(2.27)
\end{equation}
%\end{indentation}
%\begin{indentation}{0pt}{0pt}{0pt}\end{indentation}
where $x=r\sqrt{|G|}$ and ${A_{n,k}^{ij} }$ are normalization 
constants that depend on the physical quantities {$\omega $} and 
{$\omega $}$_{c}$. The lower 
component is obtained by substituting (\ref{eq22}) and (\ref{eq23})  into the "kinetic balance" 
relation (\ref{eq16}). Doing so while exploiting the differential and recursion 
properties of the Laguerre polynomials (see Appendix A) we obtain the 
following for $\varepsilon  > 0$:
%\begin{equation}
\beqn
%\begin{array}{c}
\label{2.26'}
{\phi_-}(r) &=&  \frac{{\sqrt{\left|G \right|}}} {{\varepsilon^+_{n,k} +m }}\,
x^{\left|k+{1\over 2} \right|}\,  e^{ -{1 \over 2}x^2}\nonumber\\
 &&\times \left\{ {\begin{array}{*{20}{ccc}}
{A_{n,k}^{++}}\, x {\left[(s-1) L_n^{k}\left(x^2\right) + 2L_n^{k+1}\left(x^2\right)\right]},
~~~~~~~~~~~~~~~~~~~~~~~~~~~~~~~~ &  {k \geq 0}  \\
{A_{n,k}^{+-}}\; \left[(s-1) (n-k) L_n^{-k-1}\left(x^2\right)-(s+1) (n+1) L_{n+1}^{-k-1}\left(x^2\right)\right],  & 
{k  < 0}.  \\
\end{array}} \right. 
%\end{array}
\eeqn
%\end{equation}
On the other hand, repeating the same calculation for the upper component of 
the negative energy solution gives the function
\beqn\label{2.27'}
{\phi_+}(r) &=& \frac{{\sqrt{\left|G \right|}}} {{\varepsilon^-_{n,k} -m }}\,
x^{\left|k+{1\over 2}\right|}\;  e^{ -{1 \over 2}x^2}\nonumber\\
&& \times
 \left\{ {\begin{array}{*{20}{ccc}}
{A_{n,k}^{-+} }\, \left[(1+s) (n+k+1) L_n^{k}\left(x^2\right) +(1-s) (n+1) L_{n+1}^{k}\left(x^2\right)\right],  & {k  \geq 0}  \\
A_{n,k}^{--}\, x {\left[(1+s) L_n^{-k-1}\left(x^2\right) - 2L_n^{-k}\left(x^2\right)\right]},
 ~~~~~~~~~~~~~~~~~~~~~~~~~~~&  {k  < 0}  \\
\end{array}} \right.
\eeqn
which could have also been obtained by applying the $\cal{CPT}$ map (\ref{eq18}) to (\ref{2.26'}). 
Thus, the structure of the whole Hilbert space solution consists of eight disconnected 
spaces that could be displayed in tabular form as shown in Table 1:\\
\begin{center}
\begin{tabular}{|c|c|c|c|c|c|c|c|}
\hline
%\bf \ta Energy \ta  &\bf Azimuthal number &\bf  Normalization\\[4mm] \hline
%\ta 1989-1993 %\ta     
{\bf Frequency} & $\om  > \om_c$  &  $\om  > \om_c$  & $\om  < \om_c $ & $\om  < \om_c $\\[4mm] 
\hline 
%$"\qquad "$ 
{\bf Energy} & $\varepsilon  > 0$
&  $\varepsilon  < 0$ &  $\varepsilon  > 0$ &  $\varepsilon  < 0$\ \\[4mm] \hline
{\bf Azimuth} & $ k\geq  0$  \qquad $k <  0$  &   $ k\geq  0$  \qquad $k <  0$   & $ k\geq  0$ \qquad $k <  0$ 
&  $ k\geq  0$ \qquad  $k <  0$        \\[4mm] \hline
\end{tabular}\\
\end{center}
\vspace{1mm}
\begin{center}
\tablename{1: Full space solution.}
\end{center}
Using the standard definition, we calculate all involved normalization constants in the above
wavefunctions. These are summarized
in the Table 2, where as stated above $s=\mbox{sign}(G)=|G|/G$:\\

\begin{center}
\begin{tabular}{|c|c|c|}
\hline
\bf  Energy   &\bf Azimuth &\bf  Normalization\\[4mm] \hline
%\ta 1989-1993 %\ta     
$\varepsilon  > 0$
& $k \geq 0$ & ~~ $A_{n,k}^{++} = \sqrt{2n! \left\{ \pi (n+k)!
\left[1+{4{|G|}\over (\varepsilon^+_{n,k}+m)^2} \left\{ 2(n+k+1) +
 n({{1-s}})\right\}\right]\right\}^{-1}}$\ \\[6mm] \hline 
%$"\qquad "$ 
 $\varepsilon  > 0$
&  $k <  0$ & ~~~ $A_{n,k}^{+-} = \sqrt{n!\left\{\pi (n-k)!
\left[1+ {2{|G|}\over (\varepsilon^+_{n,k}+m)^2} 
\left\{2(n+1) + (k+1) (s-1)\right\}\right]\right\}^{-1}}$ \ \\[6mm] \hline
%\ta             \ta    
$\varepsilon <  0$ &     $k \geq 0$ &\   $A_{n,k}^{-+} = \sqrt{n!\left\{ \pi (n+k+1)!
\left[1+{2{|G|}\over (\varepsilon^-_{n,k}-m)^2} \left\{ 2(n+1) + k(s+1) 
\right\}\right]\right\}^{-1}}$           \\[6mm] \hline
%\ta  \ta    
%$"\qquad "$
 $\varepsilon <  0$
&  $k <  0$  &~~~       $A_{n,k}^{--} = \sqrt{2n!\left\{ \pi (n-k-1)!
\left[1+ {4{|G|}\over (\varepsilon^-_{n,k}-m)^2} \left\{ 2(n-k) + n(s+1)\right\}\right]\right\}^{-1}}$        \\[6mm] \hline
\end{tabular}
\end{center}
\vspace{1mm}
\begin{center}
\tablename{2: Normalization in terms of different physical quantities.}
\end{center}

%%%%%%%%%%%%%%%%%%%%%%%%%%%%%%%%%%%%%%%%%%%%%%%%%%%%%%%
\section{Discussions}
%%%%%%%%%%%%%%%%%%%%%%%%%%%%%%%%%%%%%%%%%%%%%%%%%%%%%%%

It is worthwhile investigating the basic features
of some limits of our
results and their interesting underlying properties. 
We consider three different cases corresponding to the 
relative strength  of the
external magnetic field (cyclotron freqency) 
to the oscillator coupling (oscillator frequency).
We also demonstrate the added value of our results as opposed of others 
in literature,  in particular the classic work of  Villalba 
and Maggiolo~\cite{villalba}.

%%%%%%%%%%%%%%%%%%%%%%%%%%%%%%%%%%%%%%%%%%%%%%%%%%%%%%%
\subsection{Energy spectrum properties}
%%%%%%%%%%%%%%%%%%%%%%%%%%%%%%%%%%%%%%%%%%%%%%%%%%%%%%%

To investigate the underlying symmetry of the system, one may study the properties 
of quantum numbers pairs $(n,k)$. However, these may not provide simple hints on 
the ordering of the energy eigenvalues $\varepsilon _{n,k}^{\pm}$, with the exception 
of two limiting cases: the weak and strong field. 

%\begin{enumerate}
%%%%%%%%%%%%%%%%%%%%%%%%%%%%%%%%%%%%%%%%%%%%%%%%%%%%%%%
\subsubsection{Weak field case}
%%%%%%%%%%%%%%%%%%%%%%%%%%%%%%%%%%%%%%%%%%%%%%%%%%%%%%%
%\item {\bf Weak field case} \\
Suppose that the cyclotron frequency is much smaller than the oscillator frequency. That is,
$\om_c \ll \om$, $G\approx -m\om$ or $s=-1$. Thus, one obtains the following positive 
\begin{equation}\label{2.22aw}
\varepsilon _{n,k}^{+} |_{\om_c \ll \om} \approx  m\sqrt{1 + {2\om\over m} 
\left[2n  +{k} (s'-1)\right]}
\end{equation}
and negative energy spectrum
\begin{equation}\label{2.22bw}
\varepsilon _{n,k}^{-}|_{\om_c \ll \om}  \approx - m\sqrt{1 + {2\om\over m} 
\left[2n +1+s' + k(s'-1)\right]}.
\end{equation}
Consequently, for $k\geq 0$ (i.e. $s'=+1$) the energy spectrum is (semi-) infinitely 
degenerate since it becomes independent of $k$. Moreover, the two spectra are 
related as 
\beq
\varepsilon_{n,k}^{-}|_{\om_c \ll \om} = -\varepsilon_{n+1,k+1}^{+}|_{\om_c \ll \om}. 
\eeq
However, for 
$k<0$  $(s'=-1)$ we obtain
\begin{equation}\label{2.22aw2}
\varepsilon _{n,k}^{+} |_{\om_c \ll \om} \approx  m\sqrt{1 + {4\om\over m}(n-k)}
= - \varepsilon _{n,k}^{-}|_{\om_c \ll \om}.
%\varepsilon _{n}^{-}|_{\om_c \ll \om_0}  \approx - m\sqrt{1 + {4\om\over m} 
%\left(n +{1}\right)}
\end{equation}
It is also interesting to note that for $k\geq 0$ there exits a 
positive energy zero mode corresponding to  $\varepsilon _{0,0}^{+}|_{\om_c \ll \om}=m$ 
with the following spinor wavefunction:
\begin{equation}
{\phi_0}(r,\te) = {A_{k}^{+}\over \sqrt{r}} \, e^{-{i\over 2}\si_3\te}\;
\left(\sqrt{m\om}\; r e^{i\te}\right)^{k+{1\over 2}}
 \exp\left[{-{1\over 2}m\om r^2}\right]
%{1\choose 0}
 \left( {\begin{array}{*{20}{c}}
1  \\
0  \\
\end{array}} \right)
\end{equation}
where the normalization:
\beq
A_{k}^{+} = A_{0k}^{++} = \sqrt{2\left\{\pi k! [1+2{\om \over m}(k+1)]\right\}^{-1}}.
\eeq
These results are in good agreement with those of Dirac fermions in the plane 
in the presence of a constant perpendicular magnetic field. 

%\item {\bf Strong field case}\\
%%%%%%%%%%%%%%%%%%%%%%%%%%%%%%%%%%%%%%
\subsubsection{Strong field case}
%%%%%%%%%%%%%%%%%%%%%%%%%%%%%%%%%%%%
Now, if the cyclotron frequency is much larger than the oscillator frequency then 
$G\approx m\om_c$
 and we obtain the positive relativistic energy spectrum:
 \begin{equation}\label{2.22as}
\varepsilon _{n,k}^{+} |_{\om_c \gg \om} \approx  m\sqrt{1 + {2\om_c\over m} 
\left[2(n +{1}) +{k} (1+s')\right]}
\end{equation}
as well as the negative one:
\begin{equation}\label{2.22bs}
\varepsilon _{n,k}^{-}|_{\om_c \gg \om}  \approx - m\sqrt{1 + {2\om_c\over m} 
\left[2n + (k+1)(1+s')\right]}.
\end{equation}
They are related to each other as
\begin{equation}
 \varepsilon_{n,k}^{+}|_{\om_c \gg \om}= \varepsilon_{n+1,k-1}^{-}|_{\om_c \gg \om}.
\end{equation}
Thus, in this case the infinite degeneracy of the spectrum corresponds to negative 
values of the azimuthal quantum number (i.e., $s'=-1$) where
$\varepsilon_{n}^{+}= \varepsilon_{n+1}^{-}$.
Here, a negative energy zero mode 
exits that corresponding to $\varepsilon_{0}^{-}|_{\om_c \gg \om}= -m$ with the 
following spinor wavefunction:
\begin{equation}
{\phi_0}(r,\te) = {A_{k}^{-}\over \sqrt{r}}\; e^{-{i\over 2}\si_3\te}\;
\left(\sqrt{m\om_c} \;r e^{i\te}\right)^{-k-{1\over 2}} 
\exp\left[{-{1\over 2}m\om_c r^2}\right]
\left( {\begin{array}{*{20}{c}}
0  \\
1  \\
\end{array}} \right)
\end{equation}
where the normalization: 
\beq
A_{k}^{-} = A_{0k}^{--} = \sqrt{2\left\{\pi (-k-1)! [1-2{\om_c \over m}k]\right\}^{-1}}.
\eeq
Comparing the week and strong magnetic field limits one can conclude that the dominant frequency 
that controls the physics of the problem is interchanged between the oscillator 
and the magnetic field as $\om \leftrightarrow\om_c$. More precisely, 
the $k$-independent infinitely degenerate energy spectra are related 
to each other as follows:
\begin{equation}\label{2.22as2}
\varepsilon _{n}^{+} |_{\om_c \ll \om, k<0} = \varepsilon _{n+1}^{+} |_{\om_c \gg \om, k\geq 0},
\qquad
\varepsilon _{n}^{-}|_{\om_c \ll \om, k\geq 0} = \varepsilon _{n+1}^{-}|_{\om_c \gg \om, k<0}
\end{equation}
where
the quantum number $n$ corresponds to the Landau level index. 
The existence of a zero-mode energy is now very clear for positive (negative) energy with
$k\geq 0$ $ (k<0)$, respectively.

%%%%%%%%%%%%%%%%%%%%%%%%%%%%%%%%%%%%%%%%%%
\subsubsection{Fine tuned case}
%%%%%%%%%%%%%%%%%%%%%%%%%%%%%%%%%%%%%%%%%%

If the oscillator frequency is tuned to resonate with the cyclotron 
frequency (i.e., $\om\approx \om_c$) 
then $G=m\Delta$, where $\Delta = \om_c -\om$ 
such that $|\Delta|\ll m$. 
In this case, the relativistic energy spectrum approaches the nonrelativistic 
energy limit 
\beq
E= {1\over 2m} (\varepsilon^2 -m^2)
\eeq
 giving the quantity:
\begin{equation}
 E^{\pm}_{nk}= |\Delta| \left[2n+1 + {s+s'\over 2} \pm {s-s'\over 2}
+k(s+s')\right].
\end{equation}
In this case, the energy spectrum degeneracy occurs whence the quantum numbers 
associated with the two states  $\psi_1$ and $\psi_2$ satisfy
the relation
\begin{equation}
 {n_2 - n_1\over k_2 - k_1} =-{s+s'\over 2}.
\end{equation}
That is, when the ratio of the shift in the principal quantum number 
is matched with the shift in the azimuthal number either up or down 
depending on the relative strength of the two frequencies and sign of $k$.

%%%%%%%%%%%%%%%%%%%%%%%%%%%%%%%%%%%%%%%%%%%%%%%%
\subsection{Comparisons with other studies}
%%%%%%%%%%%%%%%%%%%%%%%%%%%%%%%%%%%%%%%%%%%%%%%

We compare our results with those in very similar studies found elsewhere 
in the literature. Such as the classic work by Villalba and Maggiolo~\cite{villalba} 
and, in particular, the energy spectrum and spinor wavefunction.
As for the latter, our is identified with theirs according to
\begin{equation}
\left( {\begin{array}{*{20}{c}}
 \phi_+ \\
 \phi_-\\
\end{array}} \right)
 \equiv 
\left( {\begin{array}{*{20}{c}}
 \psi_1 \\
 i \psi_2\\
\end{array}} \right).
\end{equation}
In what follows, we summarize our remarks regarding few points in~\cite{villalba}: 
\begin{enumerate}
\item The imaginary $i$ is missing from some of the off diagonal 
entries in the Dirac equation (11), however, it was 
later corrected in (28) and (29).
\item The relative strength of the cyclotron frequency $\omega_{c}$ to 
the oscillator frequency $\omega $
(i.e., whether {$\omega $}$_{c}$ is greater than or less than {$\omega $}) is ignored.
\item In addition, the negative energy solutions was also ignored altogether. Thus, only 
one fourth of the regular solution space,
which consists of eight subspaces and whose structure is shown in Table 1, 
 was obtained in~\cite{villalba}. The authors obtained only the two subspaces corresponding
to  $\varepsilon > 0$  and $\om_c> \om$.
\item The alternative signs in (31), which correspond to the 
sign of the energy, was confused with the independent signs for the physical parameter 
{$\mu $} ($k$ in our notation). % obtained in (34).
\item The relative number of nodes of the top to the bottom spinor components 
for $k < 0$ as given by equations (\ref{2.26'}) and (\ref{2.27'}) is incompatible with the
"kinetic balance" relation (\ref{eq16}).
\end{enumerate}

%%%%%%%%%%%%%%%%%%%%%%%%%%%%%%%%%%%%%%%%%%%%%%%%%%%%%%%%%%%%%%%%%%%%%
\section{Density of current}
%%%%%%%%%%%%%%%%%%%%%%%%%%%%%%%%%%%%%%%%%%%%%%%%%%%%%%%%%%%%%%%%%%%%%

We examine the behavior of the present system by analyzing the
electric current density. Indeed, from our findings we can show 
\begin{equation}
 %{J= \left [\Phi^* \si_x \Phi,  \Phi^* \si_y \Phi\right]
\vec J \sim \langle \vec \al\rangle = i\langle  \si_3 \vec \si\rangle.
\end{equation}
For this calculation, we use the spinor wavefunction obtained above. 
This
 gives a null value in Cartesian coordinates, which is
$J_x=J_y=0$. 
This, of course, is expected since there is no net charge drift. 
As a reassuring exercise, we calculate the same current
in cylindrical coordinates
\begin{equation}\label{rcurr}
 J_r = \vec J \cdot \hat r = i\langle  \si_3\vec \si \cdot \hat r\rangle,
\qquad
 J_{\theta} = \vec J \cdot \hat {\theta} = i\langle  \si_3\vec \si \cdot \hat \theta\rangle.
\end{equation}
In this calculation, we employ the similarity transformation (\ref{eq8}). 
The calculation gives $J_r=0$; however $J_\te$ does not vanish having 
the components given in Table 3. This is due to the fact that 
the physical problem in cylindrical coordinates is for a charged particle 
confined to a circular motion due to the constant magnetic field. \\ 
\begin{center}
\begin{tabular}{|c|c|c|c|c|c|}
\hline
\bf  Energy  &\bf Azimuth  &\bf    Angular Current Component  \\[2mm] \hline 
%\ta 1989-1993 %\ta     
 $\varepsilon  > 0$%\vline
& $k \geq 0$ &  $J_{\theta}^{++} = {8\sqrt{|G|}\over \varepsilon^+_{n,k}+m}\
\left[n+ (k+1) {{s+1}\over 2}\right] \left\{ 1+{4|G|\over (\varepsilon^+_{n,k}+m)^2 }
\left[ 2(n+k+1) + n{{(1-s)}}\right] \right\}^{-1}$\ \\[5mm] \hline 
 $\varepsilon  > 0$
&  $k <  0$ & $ J_{\theta}^{+-} = {2\sqrt{|G|}\over \varepsilon^+_{n,k}+m}\
(k+1){{s-1}\over 2} \left\{ 
1+{4|G|\over (\varepsilon^+_{n,k}+m)^2} \left[ 2(n+k+1) + n(1-s)\right]\right\}^{-1}$~~~~~~~~ \ \\[5mm] \hline
%\ta             \ta    
 $\varepsilon <  0$ &     $k \geq 0$ & $J_{\theta}^{-+} = -{2\sqrt{|G|}\over \varepsilon^-_{n,k}-m}\
k{s+1\over 2} \left\{
1+ {4 |G|\over (\varepsilon^-_{n,k}-m)^2} \left[ 2(n-k)+ n(1+s) \right]\right\}^{-1}$~~~~~~~~~~~~~~~~~~~
           \\[5mm] \hline
%\ta  \ta    
%$"\qquad "$ 
 $\varepsilon <  0$
&  $k <  0$  & $J_{\theta}^{--} = - {8\sqrt{|G|}\over \varepsilon^-_{n,k}-m}\
\left(n+ k {s-1\over 2}\right) \left\{ 
1+ {4|G|\over (\varepsilon^-_{n,k}-m)^2} \left[ 2(n-k) +n(1+s)\right]\right\}^{-1}$~~~~~~~~~~        
\\[5mm] \hline
\end{tabular}
\end{center}
\vspace{3mm}
\begin{center}
\tablename{3: Density of current for $4$ subspaces.}
\end{center}
One could make a different analysis in terms of the physical quantities corresponding to 
different signs $(s=\pm)$ and for all the eight different subspaces. All these analysis 
could be used to give an interesting description for the anomalous QHE.

%%%%%%%%%%%%%%%%%%%%%%%%%%%%%%%%%%%%%%%%%%%%%%%%%%%%%%%%%%%
\section{Non-relativistic limit}
%%%%%%%%%%%%%%%%%%%%%%%%%%%%%%%%%%%%%%%%%%%%%%%%%%%%%%%%%%

It is interesting to study the non-relativistic limit of our work
to reproduce results already known in literature. This can be achieved by 
taking the limit $m\lga \infty$ in the above findings. Now in the units
$\hbar=c=1$, the non-relativistic problem has already been 
worked (see, for example,~\cite{jellal1}):
\beq
{\cal H}\Psi(r,\te)
=\left[-\frac{r^2}{2m}\left(\pa_r^2+\frac{1}{r}\pa_r +\frac{1}{r^2}\pa_{\te}^2\right)
  - i\frac{1}{2}\om_c \pa_{\te} 
  + \frac{m}{8}\om^2 r^2            
 \right]\Psi(r,\te) = E\Psi(r,\te)
\eeq
The wavefunctions 
are (where $\om=0$ and $s=+1$)
\beq
\Psi_{n,\al}(r,\te)
=(-1)^n \times \frac{1}{\sqrt{\pi} l_0} \sqrt{\frac{n!}{(n+\vert\al \vert)!}} 
\exp{\left( -\frac{r^2}{2l_0^2}\right)}\, 
\left(\frac{r}{l_0}\right)^{\vert \al \vert}\, 
L_n^{(\vert \al \vert)}\left(\frac{r^2}{l_0^2} \right)\,
 e^{i\al\te} \label{2.19}
\eeq
where $n=0, 1, 2,\cdots$ is the principal quantum number,
$\al=0, \pm 1, \pm 2 , \cdots$ the angular moment quantum number
and $l_0=\sqrt{1\over eB}$, the magnetic length.  The corresponding energy eigenvalues
are given by
\beq\label{nnen}
E_{n,\al} = \Omega\left(n+\frac{\vert \al \vert+1}{2}\right)+\frac{\om_c}{2}\al.
\eeq
where $\Omega$ is the frequency $\Omega = \sqrt{\om_c^2 + 4\om^2}$.

To  compare with the non-relativistic limit of our work,
we take the limit $m\lga \infty$ and use the well-known non-relativistic
energy formula $E=(\varepsilon^2-m^2c^2)/2m$ giving
 \begin{equation}\label{nnlim}
E^{\pm}_{n,k} = 2\om_c
\left\{ {\begin{array}{*{20}{c}}
n+k+1, &  {k \geq 0}  \\
n+{1\pm 1\over 2},  & 
{k  < 0}.  \\
\end{array}} \right. \\
\end{equation}
The $\pm$ sign for $k<0$ is a remnant of the positive/negative energy spectrum 
of the relativistic theory that is exhibited as a zero energy mode in the infinitely 
degenerate part of the spectrum.

%%%%%%%%%%%%%%%%%%%%%%%%%%%%%%%%%%%%%%%%%%%%%%%%%%%%%%%%%
\section{Conclusion}
%%%%%%%%%%%%%%%%%%%%%%%%%%%%%%%%%%%%%%%%%%%%%%%%%%%%%%

The present paper was devoted to give a complete solution to the confined Dirac 
fermion system in the presence of a perpendicular magnetic field. Indeed, using a 
similarity transformation we have formulated our problem in terms of the polar 
coordinate representation that allows us to handle easily the "kinetic balance" 
relation between the two spinor components. One spinor component was obtained 
by solving a second order differential equation while the other component was 
obtained using the kinetic "balance" relation. It resulted in a full solution 
space made of 8 subspaces, which suggests that it is necessary to include all 
components of this subspace in the computations of any physical quantity. A failure 
to do so will result in erroneous conclusions. 

Our results were employed to 
discuss few important limiting cases: the weak, strong and fine tuned magnetic 
field cases. In particular, we showed that there is a symmetry between 
the negative and positive energy solutions. In the weak magnetic field case 
the system was shown to behave like a two dimensional Dirac system in the 
presence of an effective magnetic field controlled by the oscillator frequency $\om$. 
To support our analysis we compared our findings favorable with those available in the 
literature and underlined the reason behind some of our differences.

On the other hand, we analyzed the transport properties of the present system 
in terms of the current density. As expected, we found a null current in Cartesian 
coordinates, however, in polar coordinates the angular component of the current 
was non-vanishing. Finally, we studied the non-relativistic limit where known results 
were recovered. 

The emergence of the quantum Hall effect in graphene~\cite{novoselov,zhang}   
opened a good opportunity not only for experimentalists but also for theorists as well. 
Because of the relativistic nature of the fermions in grapheme and due to some additional 
constraints, the appropriate mathematical system seem to be the massless Dirac fermions. 
However, our present work suggests that present theoretical investigations in 
the literature did not include adequately contributions from all solution parameter space~\cite{jellal2}  
 and hence will lead to incomplete, and sometime erroneous, results. Extending our present 
analysis to the massless Dirac fermion system will be desirable to put the theoretical approach 
to grapheme systems on firm grounds.

Finally, we think it will be appropriate to look for the irregular solutions 
of the present problem. The importance of this issue comes from the fact that 
it will help us to construct the two-point Green function which is very much need 
in the calculation of many physical quantities and will enable us build the corresponding 
conformal theory.

%%%%%%%%%%%%%%%%%%%%%%%%%%%%%%%%%%%%%%%%%%%%%%%%%%%%%%%%%%%%%%%%%%%%%%
\section*{{Appendix A: Properties of the associated Laguerre polynomials}}
%%%%%%%%%%%%%%%%%%%%%%%%%%%%%%%%%%%%%%%%%%%%%%%%%%%%%%%%%%%%%%%%%%%%%%

 The following are useful formulae and relations satisfied
by the generalized orthogonal Laguerre polynomials $L_n^\nu
(x)$ that are relevant to the developments carried out in this
work. They are found in most textbooks on orthogonal polynomials
\cite{ref12}. We list them here for ease of reference.

The differential equation:
\begin{equation}\label{A1}
\left[ {x\frac{{{d^2}}}{{d{x^2}}} + \left(
{\nu  + 1 - x} \right)\frac{d}{{dx}} + n} \right]L_n^\nu (x) =
0
\end{equation}
where 
$x \ge 0$, $\nu  >  - 1$
and $n = 0,1,2,\cdots $. They could be expressed in
terms of the confluent hypergeometric function as
\begin{equation}\label{A2}
L_n^\nu (x) = {{}{{\Gamma (n + \nu  +
1)} \over {\Gamma (n + 1)\Gamma (\nu  + 1)}}}{}_1{F_1}( - n;\nu  +
1;x).
\end{equation}
The associated three-term recursion relation:
\begin{equation}\label{A3}
xL_n^\nu  = (2n + \nu  + 1)L_n^\nu  - (n +
\nu )L_{n - 1}^\nu  - (n + 1)L_{n + 1}^\nu
\end{equation}
Other useful recurrence relations:
\begin{equation}\label{A4}
xL_n^\nu  = (n + \nu )L_n^{\nu  - 1} - (n +
1)L_{n + 1}^{\nu  -
1}
\end{equation}
\begin{equation}\label{A5}
L_n^\nu  = L_n^{\nu  + 1} - L_{n - 1}^{\nu  +
1}
\end{equation}
The differential formula:
\begin{equation}\label{A6}
x\frac{d}{{dx}}L_n^\nu  = nL_n^\nu  - (n +
\nu )L_{n - 1}^\nu
\end{equation}
The orthogonality relation:
\begin{equation}\label{A7}
\int\limits_0^\infty  {{\rho ^\nu }(x)L_n^\nu
(x)L_m^\nu (x)dx}  = {{}{{\Gamma (n + \nu  + 1)} \over {\Gamma
(n + 1)}}}{\delta_{nm}}
\end{equation}
where 
\begin{equation}
{\rho ^\nu }(x) = {x^\nu }{e^{ - x}}.
\end{equation}

%%%%%%%%%%%%%%%%%%%%%%%%%%%%%%%%%%%%%%%%%
\section*{Acknowledgments}
%%%%%%%%%%%%%%%%%%%%%%%%%%%%%%%%%%%%%%%%%%%

The authors acknowledge the support provided by the Physics Department at 
King Fahd University of Petroleum $\&$ Minerals under project FT-090001. We are 
also grateful to the Saudi Center for Theoretical Physics (SCTP) for the 
generous support.


\begin{thebibliography}{1}
\bibitem{ref1}
%[1]
T. Chakraborty, {\it Comments Cond. Mat. Phys.} {\bf 16} (1992) 35.

\bibitem{prange}
%[2]
R.E. Prange and S.M. Girvin, editors, "The Quantum Hall Effect",
(Springer, New York 1990).

\bibitem{ref3} L.D. Landau and E.M. Lifschitz, "{Quantum Mechanics}",
(Pergamon, New York, 3$^{rd}$ edition 1977).

\bibitem{ref4} C. Kittel, "Introduction to Solid State Physics", (John Wiley \& Sons,
New York 1986).

\bibitem{ando} Y. Zheng and T. Ando, \PR {\bf B65} (2002) 245420.

\bibitem{sharapov} V.P. Gusynin and S.G. Sharapov, \PRL {\bf 95}
(2005) 146801.
%\bibitem{ref5}
\bibitem{novoselov} K.S. Novoselov, A.K. Greim, S.V. Morosov, D. Jiang,
M.I. Katsnelson, V.I. Grigorieva, L. Levy, S.V. Dubonos and A.A. Firsov,
{\it Nature} {\bf 438} (2005) 197.

\bibitem{zhang} Y. Zhang, Y.-W. Tan, H.L. St\"ormer and P. Kim,
 {\it Nature} {\bf 438} (2005) 201.
 

\bibitem{ref6} A.M.J. Schakel, {\it Phys. Rev.} {\bf D43} (1991) 1428; A. Neagu and A.M.J. 
Schakel, {\it Phys. Rev.} {\bf D 48} (1993) 1785.

\bibitem{goerbig}
M.O. Goerbig and N. Regnault, \PR {\bf B74} (2006 ) {161407}.

\bibitem{toke}C. T\"oke, P.E. Lammert, J.K. Jain and V.H. Crespi,
\PR {\bf B74} (2006) {235417}.

\bibitem{khvesh} D.V. Khveshchenko, \PR {\bf B75} (2007) {153405}.

\bibitem{toke2} C. T\"oke and J.K. Jain, \PR{\bf B75} (2007) 244540,
{\sf cond-mat/0701026}.

%\bibitem{goerbig2} M.O. Goerbig and N. Regnault,
%{\it Phys. Rev.} {\bf B75} (2007) 241405, {\sf cond-mat/0701661}.
%\bibitem{prange} For instance see R.E. Prange and S.M. Girvin (editors), "The Quantum Hall
%Effect", (Springer, New York 1990).

\bibitem{jellal2} A. Jellal, \NP {\bf B804} (2008) 361, {\sf hep-th/0505095}.

\bibitem{villalba} V.M. Villalba and A.A.R. Maggiolo, {\it Eur. J. Phys.} {\bf B22} (2001) 31, 
{\sf cond-mat/0107529}.


\bibitem{jellal1} J.P. Gazeau, P.Y. Hsiao and A. Jellal, \PR{\bf B65} (2002) 094427, 
{\sf cond-mat/0101338}.

\bibitem{alhaidari} A.D. Alhaidari, {\it Ann. Phys.} {\bf 320}  (2005) 453.

\bibitem{explain} Any other choice for the pair of Pauli matrices can be obtained 
from the present one through a unitary transformation, hence leaving 
the physics of the problem unaltered.

\bibitem{ref12} I.S. Gradshteyn and I.M. Ryzhik, "Table of Integrals, Series, And 
Products", (Academic Press, New York 1980). 




\end{thebibliography}
\end{document}